\documentclass{aastex}
\usepackage{emulateapj5,apjfonts,epsfig}

\slugcomment{To appear in the Astrophysical Journal Letters}

\shorttitle{NEAR-IR COUNTERPART TO AXP 1E 1048.1$-$5937}
\shortauthors{WANG \& CHAKRABARTY}

\begin{document}
\bibliographystyle{apj_noskip}

\title{The Likely Near-Infrared Counterpart to the Anomalous X-Ray Pulsar
1E~1048.1$-$5937\altaffilmark{1}} 

\author{Zhongxiang Wang\altaffilmark{2} and Deepto
Chakrabarty\altaffilmark{2,3}} 

\affil{Department of Physics and Center for Space Research,
Massachusetts Institute of Technology, Cambridge, MA 02139;} 
\email{wangzx@space.mit.edu, deepto@space.mit.edu}

\altaffiltext{1}{Based in part on observations from the 6.5~m Baade
telescope operated by the Observatories of the Carnegie Institution of
Washington for the Magellan Consortium, a collaboration between the
Carnegie Observatories, the University of Arizona, 
Harvard University, the University of
Michigan, and the Massachusetts Institute of Technology.}  

\altaffiltext{2}{Visiting astronomer, Cerro Tololo Inter-American
Observatory, National Optical Astronomy Observatories, operated by the
Association of Universities for Research in Astronomy under contract
to the National Science Foundation.}

\altaffiltext{3}{Alfred P. Sloan Research Fellow.}

\begin{abstract}
We report our discovery of the likely near-infrared counterpart to the
anomalous X-ray pulsar (AXP) 1E~1048.1$-$5937, using observations from
the 6.5~m Baade (Magellan~I) telescope in Chile.  We derived a precise
position for the X-ray source using archival data from the {\em
Chandra X-Ray Observatory}.  This position is inconsistent with a
position reported earlier from {\em XMM-Newton}, but we show that the
originally reported {\em XMM-Newton} position suffered from attitude
reconstruction problems. Only two of the infrared objects in a
17\arcsec$\times$17\arcsec\ field containing the target have unusual
colors, and one of these has colors consistent with those of the
identified counterparts of two other AXPs.  The latter object is also
the only source detected within the 0$\farcs$6 {\em Chandra} error
circle, and we identify it as the counterpart to 1E~1048.1$-$5937.
This is the first AXP counterpart detected in multiple infrared bands,
with magnitudes $J$=21.7(3), $H$=20.8(3), and $K$=19.4(3). There is
marginal evidence for spectral flattening at longer wavelengths.
\end{abstract}


\section{INTRODUCTION}

The anomalous X-ray pulsars (AXPs) are a small group of neutron stars
with spin periods falling in a narrow range (6--12~s), very soft X-ray
spectra, and with no evidence of binary companions (see Mereghetti et
al. 2002 for a recent review).  Their X-ray luminosities greatly
exceed the power available from spin-down of the pulsars.  These
objects are thus believed to be isolated neutron stars either having
extremely strong ($\sim10^{14}$ G) surface magnetic fields
(``magnetars'') or accreting from a residual accretion disk.  X-ray
bursts detected from the AXPs 1E~1048.1$-$5937 (Gavriil, Kaspi, \&
Woods 2002) and 1E~2259+586 (Kaspi \& Gavriil 2002) have strengthened
an already suspected connection to the soft gamma-ray repeaters
(SGRs).  The recent identification of optical/infrared counterparts to
the AXPs 4U~0142+61 (Hulleman, van Kerkwijk, \& Kulkarni 2000) and
1E~2259+586 (Hulleman et al. 2001) and the discovery of optical
pulsations from 4U~0142+61 (Kern \& Martin 2002) favor the magnetar
scenario for these objects.  In this paper, we report on the detection
of a third AXP counterpart, and the first measurement of infrared
colors for one of these objects. 

The AXP 1E~1048.1$-$5937 ($l$=288\fdg3, $b$=$-$0\fdg5) was discovered
serendipitously in 1979 during an {\em Einstein} observation of the
Carina Nebula (Seward, Charles, \& Smale 1986).  It has since been
extensively observed by a series of X-ray missions including {\em
EXOSAT} (Seward et al. 1986), {\em Ginga} (Corbet \& Day 1990), {\em
ROSAT} (Mereghetti 1995), {\em ASCA} (Corbet \& Mihara 1997; Paul et
al. 2000), {\em RXTE} (Mereghetti, Israel, \& Stella 1998; Baykal et
al. 2000; Kaspi et al. 2001; Gavriil et al. 2002), and {\em
XMM-Newton} (Tiengo et al. 2002).  These observations establish the
source as a 6.4~s pulsar that is spinning down, with an absorbed
power-law + blackbody X-ray spectrum and occasional SGR-like bursts.
The high column density to the source indicates that it lies behind the
Carina Nebula, setting a lower limit on its distance of $d\gtrsim
2.8$~kpc (Seward et al. 1986; \"Ozel, Psaltis, \& Kaspi 2001).
Previous searches for an optical counterpart have been unsuccessful
(Seward et al. 1986; Mereghetti, Caraveo, \& Bignami 1992) with 
the resultant limiting magnitudes $BVR \gtrsim 24.3$ 
(Israel, Mereghetti, \& Stella 2001).

\section{OPTICAL AND INFRARED OBSERVATIONS}

We obtained optical images of the field around 1E~1048.1$-$5937 on
2001 March~24 using the Raymond and Beverly Sackler Magellan Instant
Camera (MagIC) at the west f/11 Nasmyth focus of the 6.5~m Baade
(Magellan~I) telescope at Las Campanas Observatory in Chile.  MagIC is
a 2048$\times$2048 SITe CCD with a 0\farcs069~pixel$^{-1}$ plate scale
and a 142\arcsec\ field of view.  Our observations were made using the
Sloan filter set (Fukugita et al. 1996).  We obtained 10~min exposures
in the $r'$ and $i'$ bands. The seeing during the night was around
1\arcsec.  The limiting magnitudes (3$\sigma$) reached were $r'=24.8$
and $i'=24.4$. Photometric standards GD~108 and RU~152F were observed
at various airmasses for calibration (Landolt~1992)

We also obtained $JHK_s$ near-infrared images of the 1E~1048.1$-$5937
field on 2002 February~25 using the Ohio State Infrared
Imager/Spectrometer (OSIRIS; Depoy et al. 1993) at the f/14 tip-tilt
focus of the 4~m Blanco telescope at the Cerro Tololo Inter-American
Observatory (CTIO) in Chile.  The detector in OSIRIS was a Rockwell
HAWAII HgCdTe 1024$\times$1024 array.  We used the OSIRIS f/7 camera,
which had a 0\farcs161~pixel$^{-1}$ plate scale and a 93\arcsec\ field
of view.  For each wavelength band, we obtained nine images in a
3$\times$3 grid with offsets of about 10\arcsec\ to allow for
correction of the rapidly varying infrared sky background and to
minimize the effect of \hfill bad 

\begin{center}
\includegraphics[scale=0.7]{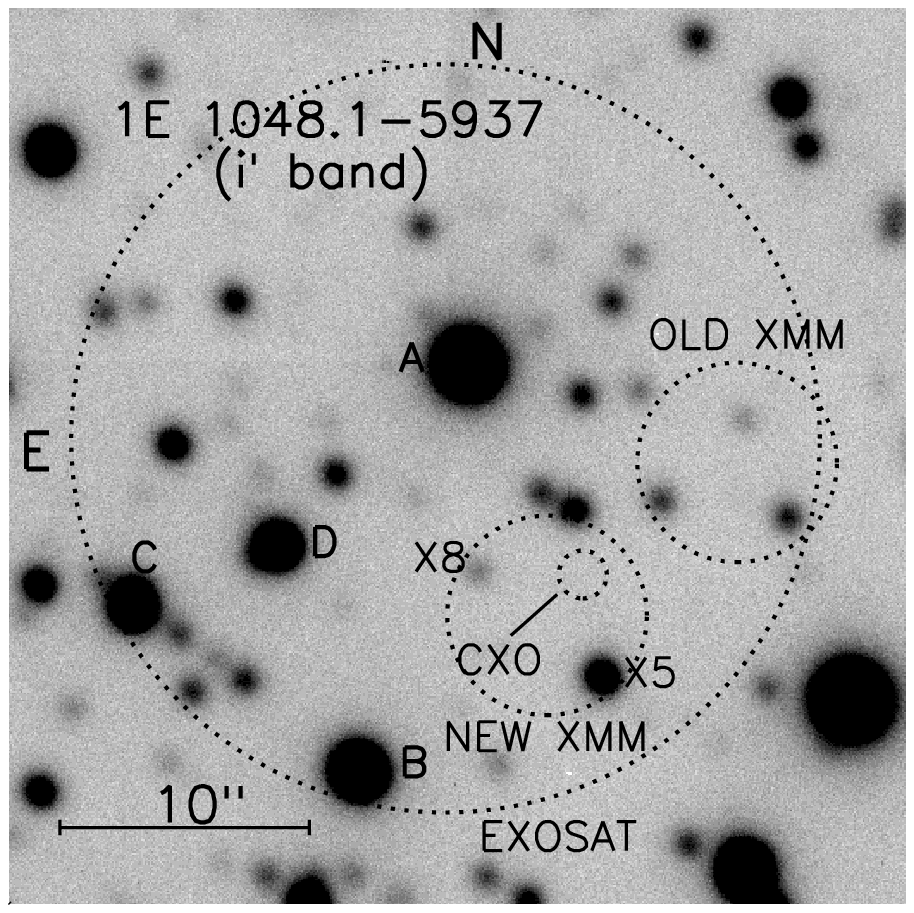}
\includegraphics[scale=0.7]{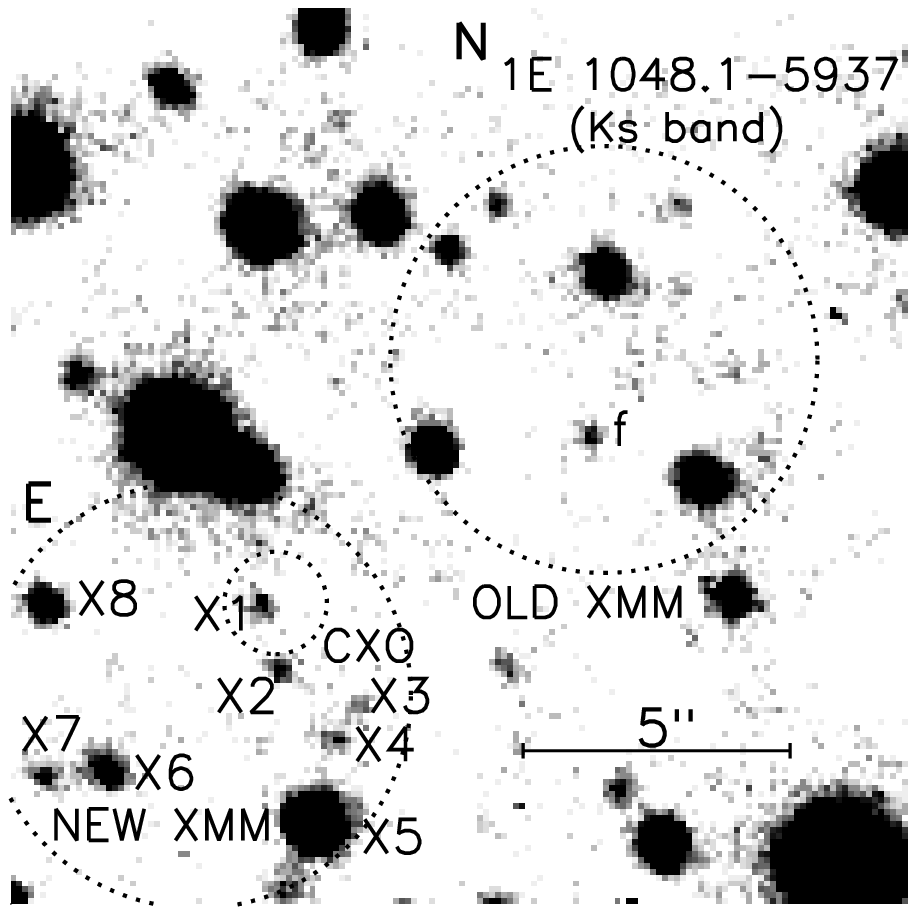}
\figcaption{Optical and near-infrared images of the 1E~1048.1$-$5937
field. {\em Top panel:} $i'$-band image. {\em Bottom panel:}
$K_s$-band Magellan image.  On both images the {\em Chandra} (CXO)
and {\em XMM-Newton} (old and new) error circles are shown. Also
shown is the 15\arcsec\ {\em EXOSAT} error circle on the optical image.  
The candidates {\em X1}--{\em X2} are discussed in \S4.  Objects $A$,
$B$, $C$, and $D$ in the optical image were studied by Mereghetti et
al. (1992).  Our proposed counterpart is object {\em X1}.}
\end{center}

\noindent
pixels.  These individual images were
shifted and combined in the course of the data reduction.  Our total
exposure in each band was 4.5~min in $J$, 9~min in $H$, and 4.5~min in
$K_s$.  During the night the seeing was good, varying from 0\farcs8 to
1\farcs2.  The limiting magnitudes (3$\sigma$) reached were $J=21.3$,
$H=20.2$, and $K_s=19.3$. The near-IR photometric standards 9136 and
LHS~2397a of Persson et al. (1998) were observed for calibration.

We obtained deeper $JHK_s$ images of the target field on 2002 April~8
with the ``Classic Cam'' near-infrared imager (Persson et al. 1992,
2001) at the west f/11 Nasmyth focus on the 6.5~m Baade telescope. The
detector was a Rockwell NICMOS3 HgCdTe 256$\times$256 array.  We
selected the camera in its low-resolution mode, with a
0\farcs112~pixel$^{-1}$ plate scale and a 29\arcsec\ field of view.
The total exposure for 1E~1048.1$-$5937 in each band was 10.5~min in
$J$, 18 min in $H$, and 15 min in $K_s$.  The same dithering and data
reduction scheme was employed as with the CTIO observations.  The
conditions were good, with the seeing around 0\farcs8.  The limiting
magnitudes (3$\sigma$) reached were $J=21.7$, $H=20.8$, and $K_s=19.7$.
The same IR standards were used as in the CTIO observations.

We used the IRAF data analysis package to reduce our data,
including the DAOPHOT routines for crowded field photometry.

\bigskip
\section{X-RAY POSITION AND ASTROMETRY}

We derived the position of 1E~1048.1$-$5937 using data from the {\em
Chandra X-Ray Observatory} public data archive. The source was
observed with the imaging detector of the {\em Chandra}
High-Resolution Camera (HRC-I; Zombeck et al. 1995) on 2001 
July~4 for 9.87 ks.  We analyzed these data using the CIAO v2.2 data
analysis package.\footnote{See http://asc.harvard.edu/ciao/}
To maximize the absolute positional accuracy of the data, we used the
CIAO thread (processing recipe) {\tt fix\_offset} to check for
systematic aspect offsets and apply the latest alignment file (2002
May 2) as necessary.  Using the CIAO tool {\tt celldetect}, we
obtained the following source position: R.A. = 10$^{\rm h}$50$^{\rm
m}$07\fs14 and Decl. = $-$59\arcdeg53\arcmin21\farcs4\ (equinox
J2000.0), with an error radius of $0\farcs6$\ (90\% confidence)
dominated by the spacecraft attitude uncertainty.   No other sources
were detected in the {\em Chandra} image.

We note that our {\em Chandra} position differs significantly
(7\farcs6) from the position reported by Tiengo et al. (2002) from
{\it XMM-Newton} observations: R.A. = 10$^{\rm h}$50$^{\rm m}$06\fs3,
Decl. = $-$59\arcdeg53\arcmin17\arcsec\ (equinox J2000.0), with an
error radius of 4\arcsec.  We reanalyzed these data and obtained a
position consistent with that of Tiengo et al. (2002).  However, an
examination of the {\em XMM-Newton} spacecraft attitude history during
this observation indicates an unusually large and oscillating offset
between the commanded and attitude-reconstructed viewing directions,
with a mean value of 10\farcs5 and an amplitude of 2\arcsec.  In
consulting with the {\em XMM-Newton} User Support Group, we learned
that such offsets usually (but not always) reflect problems with the
attitude measurement accuracy rather than a pointing error, resulting
in an erroneous reconstruction of the spacecraft attitude and the
source coordinates (M. Santos-Lleo 2002, private communication).  We
were later informed by the {\em XMM-Newton} project that they have
revised their attitude determination algorithm, placing less emphasis
on the satellite pitch angle as determined from the fine sun sensor in
the unusual case of tracking on a single guide star, as in this
observation (F. Jansen 2002, private communication).  This has led to
a revised {\em XMM-Newton} position: R.A. = 10$^{\rm h}$50$^{\rm
m}$07\fs31, Decl. = $-$59\arcdeg53\arcmin23\farcs1, equinox J2000.0,
error radius 4\arcsec.  This is consistent with the {\em Chandra}
position.  We show both the original and the revised positions in
Figure~1. We thus disregard the {\em XMM-Newton} position reported by
Tiengo et al. (2002).

An astrometric solution for the optical images was derived by matching
about 20 unsaturated field stars with the USNO-A2.0 catalog of 
astrometric standards (Monet et al. 1998). Since the field of view of our
infrared images is too small to allow a direct tie to the USNO-A2.0
system, we derived an astrometric solution for these images indirectly 
by matching stars with our optical images.  The total nominal
uncertainty for locating the {\em Chandra} \hfill source position on our
optical/infrared images is 1\arcsec

\begin{center}
\includegraphics[scale=0.7]{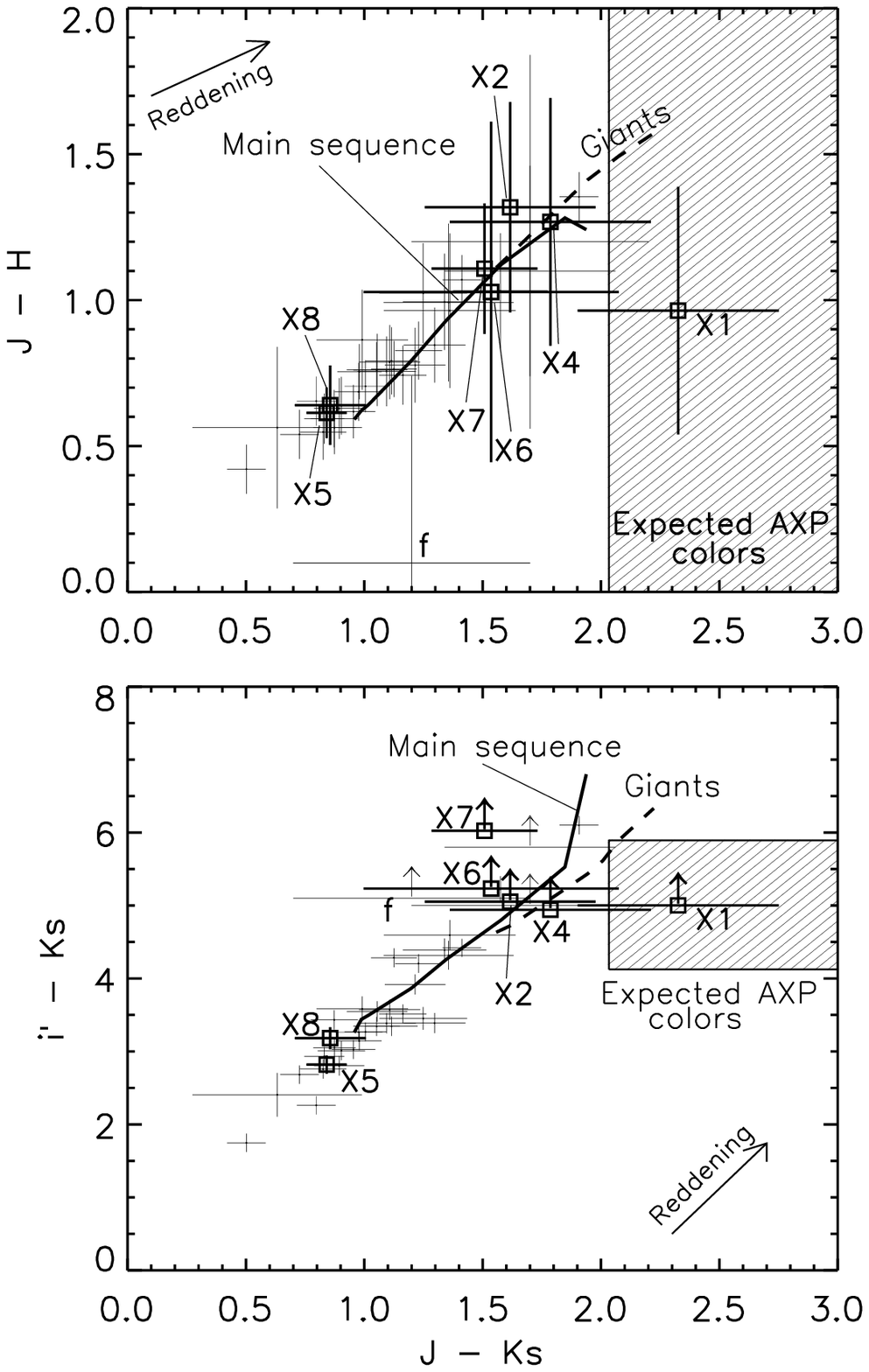}
\figcaption{Color-color diagrams for the objects in the field of
1E~1048.1$-$5937.  The dark squares indicate the candidates discussed
in \S4 and the light crosses indicate other field stars.  The
direction of increasing reddening is indicated by the arrow.  The dark
solid and dashed lines indicate the expected colors for main sequence
and giant stars, respectively, with a reddening of $A_V=5.8$.  The
shaded region indicates the expected counterpart colors given the
properties of the two other known AXP counterparts.  Only our proposed
counterpart ({\em X1}) lies in this region; it is also the only
detected source in the {\em Chandra} error circle.}
\end{center}

\noindent
(90\% confidence), which includes
the systematic uncertainty of the USNO-A2.0 catalog (0\farcs25; Monet
et al.~1998), the transformation uncertainty between the optical image
and the USNO-A2.0 catalog (0\farcs4), and the uncertainty of {\em
Chandra} observation.  For the {\em XMM-Newton} position, the quoted
4\arcsec\ uncertainty dominates.  In Figure 1, we show our $i'$ and
$K_s$ images of the field of 1E~1048.1$-$5937 with the error circles
of both {\em Chandra} and {\em XMM-Newton} plotted. 

\section{RESULTS}

As seen in the top panel of Figure~1, no objects were detected within
the {\em Chandra} error circle in our optical images, although two
objects (labeled {\em X5} and {\em X8}) were detected within the
revised {\em XMM-Newton} error circle.  Also indicated in that panel
are the candidates ($A$, $B$, $C$, and $D$) previously studied by
Mereghetti et al. (1992).  In the bottom panel of Figure~1, we
detected one object ({\em X1}) within the {\em Chandra} error circle
in our deep infrared images and another ({\em X2}) just outside.  In
addition, 

\begin{center}
\includegraphics[scale=0.7]{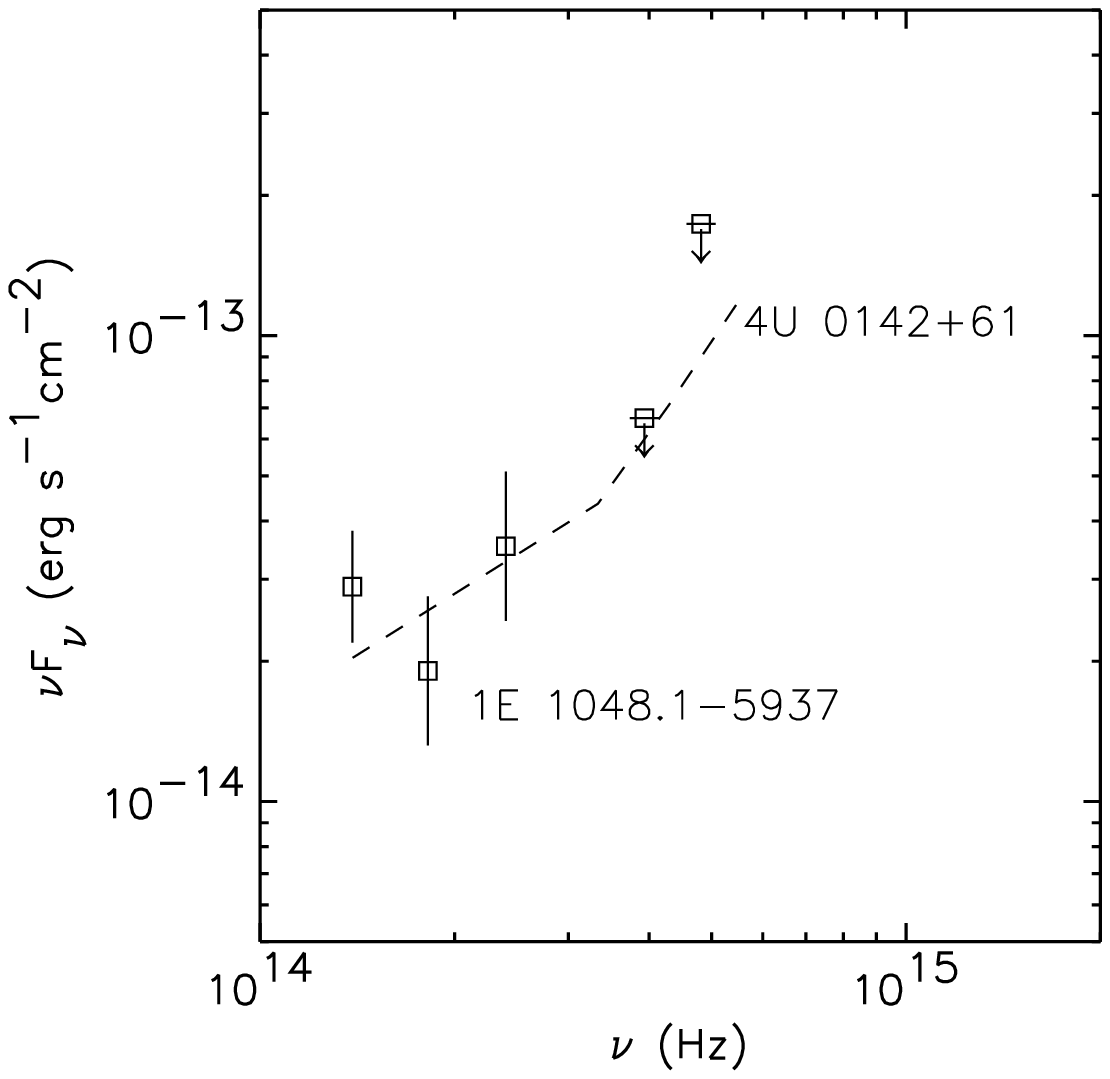}
\figcaption{Dereddened flux from object {\em X1}, the counterpart to
1E~1048.1$-$5937.  For comparison, the dashed line indicates the
interpolated dereddened flux from AXP 4U~0142+61 (Hulleman et al.~2000; 
Hulleman 2002).}
\end{center}

\noindent
we detected six other objects (labeled as {\em X3--X8})
within the revised {\em XMM-Newton} error circle.  Object {\em X3} was
detected only in the $K_s$ band; since we cannot derive colors for
this object, it is excluded from further consideration below.  The
faint objects {\em X3}, {\em X4}, and {\em X1} were not detected in
our CTIO $JHK_s$ observations, which were not as deep as the Magellan
observations.  The positions and optical/infrared magnitudes of our
candidates are summarized in Table~1.

Figure~2 shows color-color diagrams for our seven candidates as well
as all other field stars in our Classic Cam images.  We note that the
inferred hydrogen column density to 1E~1048.1$-$5937, $N_{\rm H} =
1.04(8)\times 10^{22}$ cm$^{-2}$ (Tiengo et al. 2001), implies a
visual extinction of $A_V\simeq 5.8$ for a typical dust-to-gas ratio
(Predehl \& Schmitt 1995).  For comparison in Figure~2, we also show
the expected colors for main sequence and giant stars (Bessell \&
Brett 1988; Cox 2000) reddened by $A_V=5.8$ (Rieke
\& Lebofsky 1985; Fan 1999).  The reddening vector lies roughly
parallel to this locus of points.  Most of our candidates (as well as
the other field stars) have colors consistent with those of normal
stars.  However, two objects, $f$ and $X1$, stand out with unusual colors. 
Assuming that the counterpart of 1E~1048.1$-$5937 is similar
to those of the AXPs 4U~0142+61 ($V=25.62\pm0.08$, $R=24.99\pm0.07$,
$I=23.84\pm0.06$, $K=19.6$, $A_V=5.4$; Hulleman et al. 2000; Hulleman
2002) and 1E~2259+586 ($R>26.4$, $I>25.6$, $J>23.8$, $K_s=21.7\pm
0.2$, $A_V=5.2$; Hulleman et al. 2001), we have also indicated in
Figure~2 the region of color-color space in which we would expect to
find the counterpart.  Only candidate {\em X1} has colors consistent
with these predictions.  We further note that {\em X1} is the only
object detected within the {\em Chandra} error circle.  We therefore
conclude that this object is the near-infrared counterpart to
1E~1048.1$-$5937.

\section{SUMMARY}

We have identified the likely near-infrared counterpart to 1E~1048.1$-$5937,
on the basis of coincidence with the {\em Chandra} X-ray source
position, unusual colors, and similarity to the 

\begin{figure*}[t]
\centerline{\epsfig{file=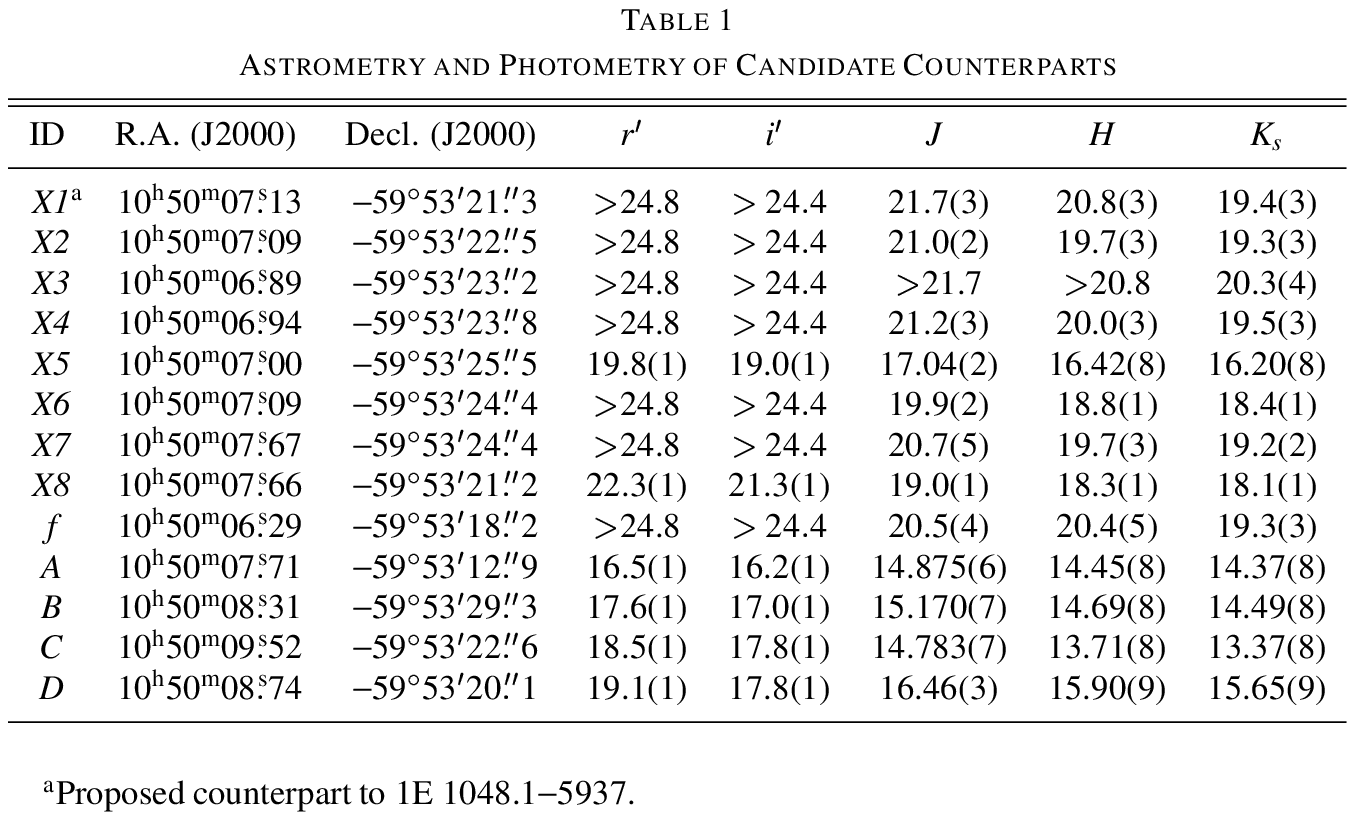}}
\end{figure*}

\noindent
two other known AXP
counterparts. We note, however, that this conclusion relies strongly on our
$i'$ non-detection.  A sufficiently faint $i'$ magnitude could
push this candidate out of the expected AXP region in the bottom
panel of Figure~2. This is the first AXP counterpart with infrared
detections in more than one band, allowing the intrinsic infrared
colors to be inferred.  Assuming $A_V=5.8$ as determined in \S4, the
dereddened infrared magnitudes of the counterpart are $J_0=20.1$,
$H_0=20.0$, and $K_{s0}=18.8$.  We plot these magnitudes and limits in
Figure~3.  For comparison, we plot the inferred spectral shape of the
4U~0142+61 counterpart (Hulleman et al.~2000; Hulleman 2002).
While there is marginal evidence for a spectral flattening
or even a turnover at the $K_s$ band, the data are also consistent
with a flat $\nu F_\nu$ infrared spectrum at the 1$\sigma$ level.  We
note that there is clear evidence for spectral flattening in the
infrared in the spectrum of 4U~0142+61 (Hulleman 2002).  It would thus be
of great interest to search for emission at longer wavelengths to 
constrain the spectral shape of AXPs further.  While there has been
relatively little theoretical work presented on the expected AXP
optical/infrared emission in the magnetar scenario, we note that
models involving X-ray illumination of a ``fallback'' accretion disk
predict a spectral turnover leading to strong emission in the infrared
and submillimeter bands (e.g., Chatterjee, Hernquist, \& Narayan 2000;
Perna, Hernquist, \& Narayan 2000).

\acknowledgments{We thank Paul Schechter for obtaining the 2001 March
Magellan data for us; Jim Elliot, Mauricio Navarrete, Hern\'{a}n
Nu\~{n}ez, Patricio Jones, Geraldo Valladanes, and the LCO staff for
their help during the 2002 April Magellan run; Eric Persson for advice
on obtaining and analyzing Classic Cam data; and Nicole van der Bliek,
Robert Blum, Angel Guerra, Sergio Pizarro, and the CTIO staff for
their assistance during our CTIO run.  We also thank Scott Burles for
useful discussions, Adrienne Juett for assistance in analyzing the
X-ray data, and Maria Santos-Lleo of the {\em XMM-Newton} User Support
Group for guidance in understanding the {\em XMM-Newton} aspect
determination problem. Finally, we thank the referee for a careful
review of our paper.}

\end{document}